\pdfoutput=1
\documentclass[a4paper,american,aip,jap,citeautoscript,floatfix,longbibliography,pdftex,superscriptaddress,%
reprint%
]{revtex4-1}
\usepackage{amsfonts,amsmath,amssymb}
\usepackage{graphicx}
\usepackage[T1]{fontenc}
\usepackage[utf8]{inputenc}
\usepackage{microtype}
\usepackage{textcomp}
\usepackage[textsize=footnotesize]{todonotes}
\usepackage{xspace}
\usepackage{hyperref}
\usepackage{hypernat}
\usepackage[todos]{CMImacros}

\newlength{\figwidth}
\newcommand{\cfeldesy}{\affiliation{Center for Free-Electron Laser Science, DESY, Notkestrasse 85,
      22607 Hamburg, Germany}}%
\newcommand{\uhhcui}{\affiliation{The Hamburg Center for Ultrafast Imaging, University of Hamburg,
      Luruper Chaussee 149, 22761 Hamburg, Germany}}%
\newcommand{\uhhphys}{\affiliation{Department of Physics, University of Hamburg, Luruper Chaussee
      149, 22761 Hamburg, Germany}}%

\begin{document}
\title{Characterizing gas flow from aerosol particle injectors}%
\author{Daniel A.\ Horke}%
\email[]{daniel.horke@cfel.de}%
\homepage[\newline]{https://www.controlled-molecule-imaging.org}%
\cfeldesy\uhhcui%
\author{Nils Roth}\cfeldesy\uhhphys%
\author{Lena Worbs}\cfeldesy%
\author{Jochen Küpper}\cfeldesy\uhhcui\uhhphys%
\date{\today}%
\begin{abstract}\noindent%
   A novel methodology for measuring gas flow from small orifices or nozzles into vacuum is
   presented. It utilizes a high-intensity femtosecond laser pulse to create a plasma within the gas
   plume produced by the nozzle, which is imaged by a microscope. Calibration of the imaging system
   allows for the extraction of absolute number densities. We show detection down to helium
   densities of $4\times10^{16}$~cm$^{-3}$ with a spatial resolution of a few micrometer. The
   technique is used to characterize the gas flow from a convergent-nozzle aerosol injector
   [Struct.\ Dyn.~2, 041717 (2015)] as used in single-particle diffractive imaging experiments at
   free-electron laser sources. Based on the measured gas-density profile we estimate the scattering
   background signal under typical operating conditions of single-particle imaging experiments and
   estimate that fewer than 50 photons per shot can be expected on the detector.
\end{abstract}
\maketitle

\section{Introduction}
\label{sec:introduction}
The advances of x-ray free-electron lasers (XFELs), which provide intense and short pulses of
coherent x-rays, open up new possibilities for imaging of aerosolized particles, and even individual
molecules, with atomic spatial resolution~\cite{Bogan:NanoLett8:310, Kuepper:PRL112:083002,
   Spence:PTRSB369:20130309, Barty:ARPC64:415}. As experiments can be conducted completely in the
gas phase and do not require sample immobilization, \eg, cryogenic freezing, XFELs furthermore
provide unprecedented capabilities for capturing ultrafast dynamics of isolated systems with
femtosecond temporal and picometer spatial resolution~\cite{Barty:ARPC64:415, Pande:Science352:725,
   Gorkhover:NatPhoton10:93, Glownia:PRL117:153003}. This is enabled by the short and intense x-ray
pulses available at these facilities, which typically provide pulses with $\sim\!1$~mJ pulse energy,
$\sim\!10$~fs pulse duration, and $\sim\!100$~pm wavelength. This allows the imaging methodology to
outrun radiation damage effects in the \emph{diffraction before destruction}
mechanism~\cite{Neutze:Nature406:752, Ziaja:NJP14:115015, Lorenz:PRE86:051911, Nass:JSR22:225}.
Combining many diffraction patterns from reproducible isolated aerosol targets imaged at different
orientations should allow one to reconstruct the three-dimensional, atomically resolved
structure~\cite{Bergh:QRB41:181, Fung:NatPhys5:64}. In recent years full 3D reconstruction has been
demonstrated and the achieved resolution continuously improved~\cite{Seibert:Nature470:78,
   Xu:NatComm5:4061, Hantke:NatPhoton8:943, Ekeberg:PRL114:098102}.

The advent of these new possibilities for imaging isolated systems \emph{in vacuo} has prompted the
development and improvement of techniques for injecting samples into the interaction region. Using
gas-dynamic virtual nozzles (GDVNs)~\cite{DePonte:JPD41:195505} for producing focused liquid jets
enabled the serial femtosecond crystallography (SFX) methodology~\cite{Chapman:Nature470:73,
   Schlichting:IUCRJ2:246}, allowing the reconstruction of sub-nanometer-resolution structures from
micrometer sized crystals~\cite{Chapman:Nature470:73, Ayyer:Nature530:202}. Aerodynamic
lenses~\cite{Liu:AST22:293, Bogan:NanoLett8:310} and convergent-nozzle
injectors~\cite{Kirian:SD2:041717} are widely used injection techniques to produce focused or
collimated streams of nano- or micrometer sized particles. They fundamentally rely on a gas flow
that interacts with the particles of interest and, through shear and drag forces, produces the
desired stream of particles. Typically, helium is used for its relatively small x-ray scattering
cross-section. However, since the helium gas density at the interaction point is still many orders
of magnitude higher than the sample density, scattering from the focusing gas can make a significant
contribution to the recorded background scattering~\cite{Kuepper:PRL112:083002, Stern:FD171:393,
   Awel:fsCXI:inprep}. In order to account for this background and to make quantitative predictions
and background calibrations, therefore, requires knowledge of the gas density at the interaction
point, typically located a few hundred micrometers below the injector tip~\cite{Kirian:SD2:041717}.

Here, we present a methodology that allows the spatially resolved measurement of gas densities down
to \mbox{$\sim4\!\times10^{16}\text{~cm}^{-3}$} with high spatial and, potentially, temporal
resolution. This is achieved by using a high-intensity femtosecond laser pulse to create a plasma
within the gas stream, which is then imaged by a microscope objective and camera. The observed
intensity of the plasma depends on the local gas pressure in the laser focus. By calibrating the
plasma formation and imaging system to known helium pressures, this method allows us to create
spatial maps of the gas flow from an injector tip. Compared to previous
methods~\cite{Golovin:AO54:3491, Landgraf:RSI82:083106}, this approach provides a higher
sensitivity, allowing the detection of one order of magnitude lower gas pressures, and it does not
rely on interferometric measurements prone to mechanical instabilities. In particular, we
characterize a convergent nozzle injector~\cite{Kirian:SD2:041717} under typical operation
conditions for XFEL single-particle diffractive imaging experiments. Based on the measured
gas-density distribution, the x-ray scattering signal expected from this helium background at
typical operating parameters of currently available XFEL endstations is calculated.

\section{Experimental Method}
\label{sec:methods}
To assess the local gas density at the tip of an aerosol injector the gas stream was crossed with a
focused Ti:Sapphire femtosecond laser beam of sufficient intensity to produce a plasma inside the
gas stream. The bright visible glow of this plasma was recorded on a camera. The intensity depended
on the laser intensity and the gas density in the interaction volume. By calibrating the imaging
system at known gas densities, this allowed us to build up a high-resolution spatial map of local
gas densities produced by the injector tip.

\begin{figure}[b]
   \centering
   \includegraphics[width=\linewidth]{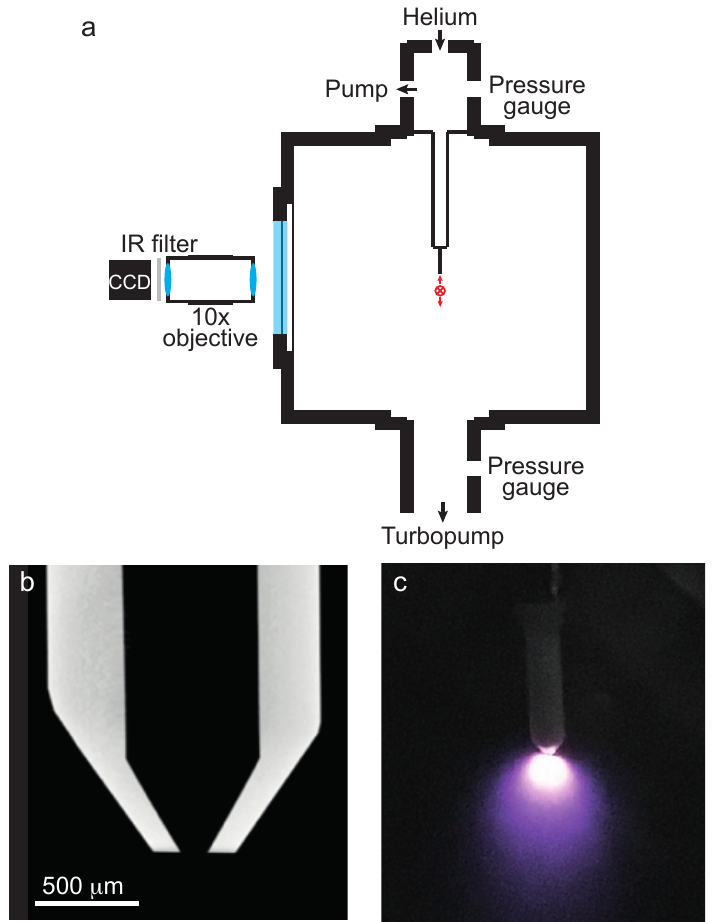}
   \caption{a) Sketch of the experimental setup and imaging system. The laser propagates out of the
      plane of the page (indicated by the red cross) and can be translated in height using a
      motorized translation stage. b) X-ray tomogram of a convergent injector tip, \cf
      reference~\onlinecite{Kirian:SD2:041717}. c) Picture of the operating injector in the vacuum
      chamber, showing the produced plasma during helium injection (recorded with a standard mobile
      phone camera).}
    \label{fig:setup}
\end{figure}
A simple sketch of the vacuum and imaging system is shown in \autoref{fig:setup}~a. The vacuum
system consisted of two differentially pumped chambers, connected only through the injector tip. The
upper chamber, \ie, upstream of the injector, contained a capacitive pressure gauge (Pfeiffer Vacuum
CMR361) with an absolute accuracy of 0.2\% independent of gas type, a high-precision leak valve
connected to a high-purity helium supply and a connection to a roughing pump, with the pumping speed
controllable through a needle valve. This setup allowed us to maintain a constant pressure during
operation of the injector by matching the helium flow into the upper chamber to the gas transmission
through the injector aperture. This chamber mimicked the typical nebulization chamber in
single-particle imaging experiments. The injector tip, with an \degree{30} convergence angle and an
111~\um orifice~\cite{Kirian:SD2:041717}, was mounted on a 6~mm outer diameter stainless steel tube
at the bottom of this upper chamber. It was located within the main vacuum chamber as shown in
\autoref{fig:setup}~a. This chamber was evacuated by a turbomolecular pump (Pfeiffer Vacuum HiPace
80) and the pressure was monitored through a full-range pressure gauge (Pfeiffer Vacuum PKR361).

The laser passed through the interaction chamber perpendicular to both, the gas-stream and the
imaging axis, as indicated by the red cross in \autoref{fig:setup}. It consisted of pulses from an
amplified Ti:Sapphire laser system (Spectra Physics Spitfire ACE) centered around 800~nm, running at
1~kHz repetition rate, and producing 40~fs pulses with 0.7~mJ per pulse used in the current
experiment. The laser beam (waist $\omega\approx5$~mm) was focused into the interaction region with
a $f=300$~mm plano-convex lens, producing a focal spot size of $50~\um$
(4$\sigma$) with a nominal Rayleigh range of $z_\text{R}\approx2.5$~mm and a peak
intensity of $8.9\times10^{14}$~W/cm$^2$. The focusing lens was placed on a 3D translation stage to
allow translation of the laser focus in space to ensure overlap with the gas stream within the
Rayleigh range and to allow probing of the local gas densities at different distances from the
injector nozzle.

The laser-matter interaction was imaged through a standard vacuum viewport with a $10\times$
long-working-distance microscope objective (Edmund Optics 59-877) that produced an image on a
high-sensitivity CMOS camera (Thorlabs DCC3240M, 10~bit monochrome, 5.3~\um pixel size). Residual
stray infrared light from the femtosecond laser was blocked using two shortpass filters (Thorlabs
FESH0700, $\text{OD}>5$ for $\lambda>700$~nm) mounted between the objective and the camera and stray
light was reduced by mechanically enclosing the optical path. The entire imaging system (objective,
filters, camera) was mounted on a three-dimensional translation stage.

The imaging system was calibrated by recording the plasma-glow intensity when flooding the chamber
to a known helium pressure; details are given in the supplementary information. To collect data from
the injector produced plasma, the injector tip was installed in the center of the chamber and the
upper chamber was pressurized with helium as discussed above. The horizontal laser-injector overlap,
\ie, along the imaging axis, was optimized to produce the brightest plasma. Then the vertical
position of the laser was adjusted by translating the focusing lens, such that it passed just below
the injector tip. The laser focus was translated downwards in steps of 12.5~\um and at every point
20~frames were collected on the camera. The exposure time was adjusted such that the plasma was
clearly visible but no saturation occurs. During the subsequent data analysis the images collected
at the same position were averaged and scaled by exposure time.

A pressure map was then produced by comparing all images taken with identical upstream pressures and
keeping for every pixel the maximum intensity value occurring in one of the images. This
``maximum-intensity-stack'' approach was chosen as the images cannot simply be averaged due to the
long-lived nature of the plasma glow. This effect is clearly visible in the photograph in
\autoref{fig:setup}~c. As the gas is moving rapidly away from the nozzle -- due to
chocked-flow conditions the speed is probably close to 1000~m/s -- glowing plasma
is observed even several millimeters below the laser excitation. Simply averaging all images
collected at different positions would therefore have overexposed the lower part of the image (since
there is intensity in this part of the image even if the excitation happens far above). Following
the combination of images, the pressure for every pixel was retrieved by comparison with the
calibration measurements. The plotted isobars were obtained from the experimental data after
applying a two-dimensional Gaussian filter with a width $\sigma=4.3$~\um.

\section{Results and Discussion}
\label{sec:results}
\begin{figure}
   \centering
   \includegraphics[width=\linewidth]{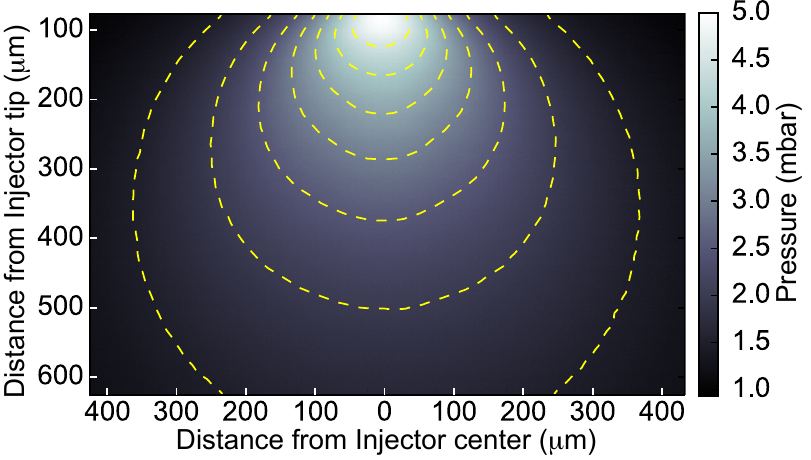}
   \caption{Pressure map recorded below the tip of a convergent injector nozzle for an upstream
      pressure of 800~mbar helium. Dashed yellow lines indicate isobars from 1.5 to 4.5~mbar in
      0.5~mbar intervals.}
   \label{fig:800mbar}
\end{figure}

The measured pressure distribution from a convergent nozzle tip operated with 800~mbar of upstream
helium is shown in \autoref{fig:800mbar}.
Similar measurements for upstream pressures of 300~mbar and 500~mbar are shown in the supplementary
materials. During the measurement the pressure in the main chamber was maintained below
$2\times10^{-2}$~mbar, ensuring chocked-flow-conditions through the orifice. The topmost measurement
was taken around 80~\um below the tip; moving the laser further up leads to clipping of the beam,
and potentially damage, on the ceramic tip. At distances $\gtrsim\!600~\um$ below the tip the
pressure had fallen such that no plasma was observed. The gas pressure was found to decrease
strongly with increasing distance from the injector tip, as expected. Due to the acceleration of gas
inside the orifice, initially some propensity for the helium to continue along the axial direction
is observed, rather than radially isotropic diffusion, resulting in the non-spherical pressure
distribution measured. Under typical operating conditions for single-particle diffractive imaging
experiments, the interaction region, that is, the crossing point of the x-ray beam with the particle
stream, is located $\sim300~\um$ below the injector tip. At this position the pressure has already
dropped considerably and, for the measurements of 800~mbar upstream pressure, shown in
\autoref{fig:800mbar}, is on the order of 3~mbar.

To quantify the spatial resolution in the produced images we differentiate between the resolution
within the imaging plane, \ie, within the plane of laser illumination, and the resolution parallel
to the camera surface. The latter is limited only by the imaging system employed. For the current
setup a single pixel corresponds to 0.86~\um (as calibrated with a microscope reticle), however we
estimate the resolution in this plane to be on the order of 2~\um due to aberrations and mechanical
instabilities. In the direction perpendicular to the imaging plane, the resolution is not only
limited by the depth of focus of the imaging system, but also by the focal spot size of the
illuminating laser, which is around 50~\um ($4\sigma$) for the data shown. This is, however, still
significantly smaller than the orifice size of the injector, allowing us to image essentially the
central slice through the (radially symmetric) pressure distribution.

\begin{figure}
   \centering
   \includegraphics[width=\linewidth]{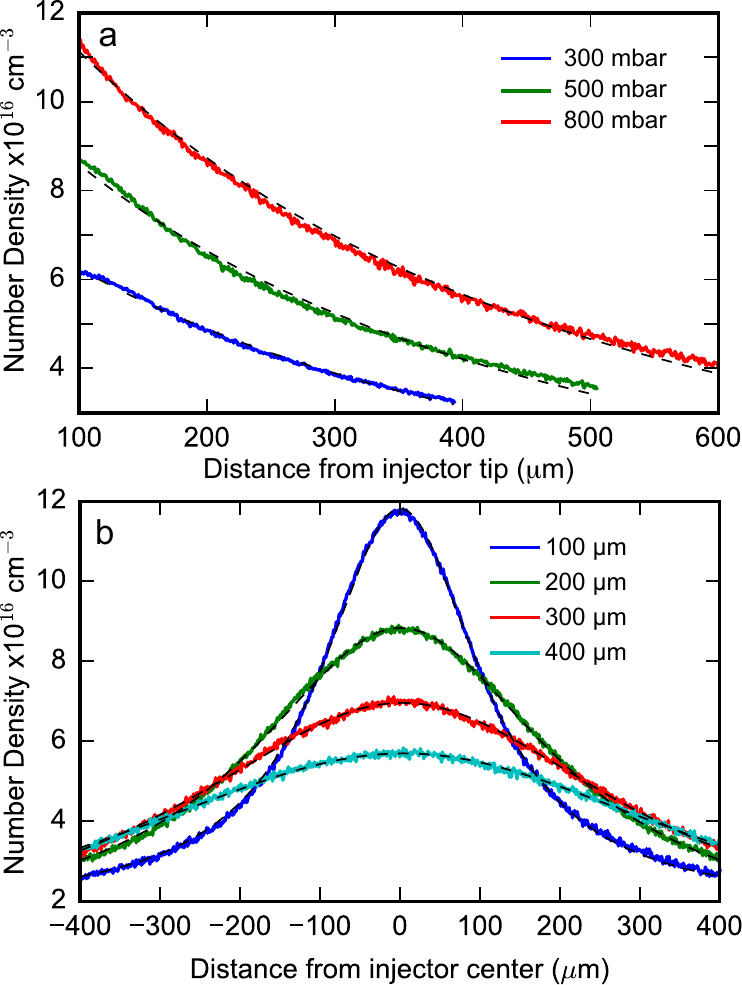}
   \caption{Gas density profiles. (a) Axial profile of the number density along the center of the
      injector as a function of distance from the tip, shown for three different upstream helium
      pressures. Dashed lines correspond to a $1/r^3$ fit. (b) Radial profiles of the number density
      across the generated plume for 800~mbar upstream pressure, at three different distances from
      the injector tip. Dashed lines correspond to a Lorentzian profile fit.}
   \label{fig:profiles}
\end{figure}

Helium pressure profiles along both the axial and radial directions are shown in
\autoref{fig:profiles}, where the measured pressure has been converted into an absolute number
density assuming ideal gas behavior.
\autoref{fig:profiles}~a shows the axial density distribution along the center line of the injector
as a function of distance from the tip, for different upstream pressures. The pressure decreases
rapidly with distance from the injector, and exhibits approximately a $1/r^3$ dependence, which is
shown by the dashed lines in \autoref{fig:profiles}~a, as would be expected for an isotropic radial
diffusion in 3D. For the production of focused nanoparticle beams the pressure upstream of the
injector is typically in the range of 200--500~mbar, while the particle focus -- and hence
interaction point -- is located a few hundred micrometer downstream the
nozzle~\cite{Kirian:SD2:041717}. Therefore, the corresponding number densities at the interaction
point are typically on the order of $5\times10^{16}~\text{cm}^{-3}$. Radial profiles of the helium
number density are shown in \autoref{fig:profiles}~b, measured at various distances below the
injector tip for an upstream pressure of 800~mbar; profiles for further upstream pressures are shown
in the supplementary information. These distributions were fit to Lorentzian functions and the good
agreement shows, that the helium gas-flow has a uniform angular distribution. These results
demonstrate that the initially narrow gas plume spreads out radially, leading to a rapid decrease in
the absolute density along the center line.

To assess the total scattering signal that can be expected from helium in XFEL based diffraction
experiments, one has to take into account not only the interaction point itself, but due to the
large Rayleigh length of the XFEL beam, typically several millimeters, one should take into account
the full extend of the helium ``cloud'' along the x-ray beam, the extend of which is visible from
the radial profiles in \autoref{fig:profiles}~b. From our spatially resolved measurements we can
assess the average helium density encountered by the XFEL pulse as it travels through the helium
cloud, and for 500~mbar upstream pressure this is $\sim3.6\times10^{16}~\text{cm}^{-3}$,
corresponding to the average helium density 300~\um below the injector tip as measured within our
field of view. Considering the known helium cross sections for elastic (Rayleigh) and inelastic
(Compton) scattering, and typical operating conditions for the CXI endstation at the Linear Coherent
Light Source (LCLS), \eg, 10~keV photon energy and $10^{11}$ photons per pulse, we expect a total of
$\sim\!500$ scattered x-ray photons per shot due to the helium background gas. Considering an
isotropic scattering distribution and a detector opening angle of \degree{60}, this corresponds to
$\sim\!40$ photons per shot on the detector. We furthermore note that the majority of these photons
($>70$~\%) originate from inelastic scattering processes, and can thus potentially be discriminated
against by an energy-resolving detector~\cite{Strueder:NIMA614:483}.

\section{Conclusion}
\label{sec:conclusion}
We present a robust and sensitive approach for measuring the spatial distribution of gas flows from
nozzles into vacuum. Calibration at known pressures allows the determination of absolute pressures
and number densities with high spatial resolution. With the current setup the minimum detectable
density is on the order of $10^{16}~\text{cm}^{-3}$, around one order of magnitude smaller than with
interferometric approaches~\cite{Golovin:AO54:3491, Landgraf:RSI82:083106}. The spatial resolution
within the imaging plane is around 2~\um, perpendicular to the imaging plane it is limited by the
laser spot size of 50~\um ($4\sigma$). We also note that this methodology can be further extended to
measurements in the time domain, due to the inherently pulsed nature of the laser illumination.

We used this approach to assess the gas flow from a convergent nozzle
injector~\cite{Kirian:SD2:041717} typically used for single-particle diffractive imaging
experiments. We found that at typical operating conditions the gas density in the interaction region
is on the order $5\times10^{16}~\text{cm}^{-3}$. By evaluating the average gas density encountered
by an x-ray pulse as it travels through the gas plume we estimate that fewer than 500~photons will
be scattered. This number could be further reduced by increasing the distance between the injector
tip and the interaction region, which could could be facilitated through the use of shallower
convergence angles within the injector~\cite{Kirian:SD2:041717}. Further approaches to reduce the
incoherent scattering from helium could incorporate inhomogeneous electric fields to deflect
particles of interest out of the helium plume~\cite{Filsinger:PCCP13:2076, Chang:IRPC34:557}, as has
been demonstrated for single molecule scattering experiments at LCLS utilizing supersonic molecular
beams~\cite{Kuepper:PRL112:083002}.

\section*{Supplementary Material}
See supplementary material for details on the calibration procedure and measured spatial profiles 
for different injector pressures.

\begin{acknowledgments}\noindent%
   We gratefully acknowledge support by Salah Awel with the experimental setup, Christian Kruse in
   an early stage of the experiment, and members of the CFEL Coherent Imaging Division for helpful
   discussions.

   In addition to DESY, this work has been supported by the excellence cluster ``The Hamburg Center
   for Ultrafast Imaging -- Structure, Dynamics and Control of Matter at the Atomic Scale'' of the
   Deutsche Forschungsgemeinschaft (CUI, DFG-EXC1074) and by the European Research Council through
   the Consolidator Grant COMOTION (ERC-Küpper-614507).
\end{acknowledgments}
\bibliography{string,cmi}
\end{document}